\documentclass[a4paper,fleqn]{cas-dc}

\usepackage[authoryear]{natbib}
\usepackage{tablefootnote}

\usepackage{listings}
\usepackage{xcolor}

\definecolor{codegray}{gray}{0.95}

\lstdefinelanguage{yaml}{
  keywords={true,false,null,y,n},
  keywordstyle=\color{blue}\bfseries,
  basicstyle=\ttfamily\small,
  sensitive=false,
  comment=[l]{\#},
  commentstyle=\color{gray}\ttfamily,
  moredelim=[l][\color{magenta}]{:},
  morestring=[b]",
  morestring=[b]'
}

\newcommand{\codename}{\texttt{UnReal-B}}
\newcommand{\codeurl}{\texttt{https://gitlab.com/neutronstars1/unreal-b/}}

\begin{document}
\let\WriteBookmarks\relax
\def\floatpagepagefraction{1}
\def\textpagefraction{.001}
\shorttitle{UnReal-B : Real-Space DFT Solver for Matter in Extreme Magnetic Fields}
\shortauthors{B. S. Pujari et~al.}

\title [mode = title]{UnReal-B : Real-Space DFT Solver for Matter in Extreme Magnetic Fields}                      


%

\author[1]{Bhalchandra S. Pujari}[orcid=0000-0002-5828-3766]
\cormark[1]
\ead{bspujari@scms.unipune.ac.in}
\ead[url]{https://scms.unipune.ac.in/~bspujari}

\credit{Conceptualization, Methodology, Software, Formal Analysis, Investigation, Writing – original draft, Writing – review \& editing}

\affiliation[1]{organization={Department of Scientific Computing Modeling and Simulation, Savitribai Phule Pune University.},
                addressline={Ganeshkhind}, 
                city={Pune},
                postcode={411007}, 
                state={Maharashtra},
                country={India}}

\author[2]{Andrey Tokarev}[orcid=0000-0003-2270-7843]
\credit{Methodology, Software, Formal Analysis, Investigation, Writing – review \& editing}
\affiliation[2]{organization={Independent Researcher},
                city={Vancouver},
                state={BC}, 
                country={Canada}}

\author[3]{Dipanjan Mitra}[orcid=0000-0002-9142-9835
   ]
\credit{Conceptualization,  Writing – review \& editing }
\affiliation[3]{organization={National Centre for Radio Astrophysics},
                addressline={Ganeshkhind}, 
                city={Pune},
                postcode={411007}, 
                state={Maharashtra}, 
                country={India}}

\cortext[cor1]{Corresponding author}

\begin{abstract}
As new observational technologies reveal the intricate details of neutron star surfaces, the demand for accessible and extensible theoretical models has never been higher. We present \codename, a real-space Density Functional Theory solver for 1D chains of matter in extreme magnetic fields ($B \approx 10^{12}$--$10^{15}$ G). By employing the adiabatic approximation to describe the spatial density, \codename\  provides a streamlined numerical framework for calculating the electronic structure of condensed atomic chains. The solver is benchmarked against existing literature for multiple astrophysically relevant elements, offering high accuracy with significantly reduced implementation overhead. \codename\ is released as an open-source package to facilitate community-driven modeling of neutron star surfaces in light of emerging observational data.
\end{abstract}

\maketitle

\section{Introduction}

Neutron stars possess the strongest magnetic fields known in the
Universe. Surface field strengths inferred from pulsar spin-down
measurements range from ($10^{8}$) to ($10^{13},\mathrm{G}$), while
magnetars exhibit dipolar fields approaching ($10^{15},\mathrm{G}$)
\citep{KaspiBeloborodov2017}. Such fields exceed the atomic unit of
magnetic field strength, ($B_0 = 2.35\times10^{9},\mathrm{G}$), by
several orders of magnitude, placing matter in a regime inaccessible
to terrestrial experiments. In this limit, electron motion
perpendicular to the magnetic field is quantized into Landau levels,
atoms become highly elongated along magnetic field lines, and chemical
bonding acquires an effectively one-dimensional character
\citep{Lai2001,Potekhin2015}. As a consequence, the outer layers of
neutron stars may consist of molecular chains and condensed phases
stabilized by strong magnetic confinement.

The microscopic properties of such matter play an increasingly
important role in neutron-star astrophysics. Measurements from NICER
have produced simultaneous mass--radius constraints with unprecedented
precision \citep{Miller2021,Riley2021,Miller2024,Riley2024}, while
IXPE has opened a new observational window through X-ray polarimetry
of strongly magnetized neutron stars \citep{Taverna2022}. Together
with multimessenger observations of neutron-star mergers
\citep{Abbott2017,Abbott2018}, these developments are transforming
neutron stars into precision laboratories for dense matter physics. At
the same time, growing evidence suggests that magnetic fields near the
stellar surface can be substantially more complex than the large-scale
dipolar geometry inferred from spin-down measurements, with localized
non-dipolar structures potentially reaching field strengths far
exceeding the global dipole component
\citep{Tiengo2013,Vigano2013,Geppert2013}. Since cohesive energies,
work functions, condensation temperatures, and ion binding energies
depend sensitively on magnetic field strength, accurate microscopic
calculations are required to connect surface composition and
condensed-matter physics to observable emission.

The importance of these quantities extends beyond thermal surface
radiation. In the partially screened gap (PSG) model of pulsar radio
emission, the supply of ions from the condensed surface regulates
particle acceleration and pair production above the polar cap
\citep{Gil2003}. The formation and stability of such gaps depend
directly on the cohesive properties of strongly magnetized matter,
which determine the ease with which ions can be extracted from the
stellar surface \citep{MedinLai2007}. Recent observational and
theoretical studies increasingly point toward a close connection
between non-dipolar magnetic fields, surface binding properties, and
the plasma conditions responsible for coherent radio emission
\citep{Mitra2024}. Accurate calculations of surface cohesive energies
and electronic structure are therefore essential for understanding
both thermal and radio manifestations of neutron stars.

The theoretical description of matter in strong magnetic fields has evolved substantially over the past five decades, driven by quantum phenomena beyond the reach of standard field-free electronic-structure methods (for a comprehensive review, see \citealt{Lai2001}). The foundational premise was established by \citet{Ruderman1971}, who argued that the surface of a highly magnetized neutron star should condense into a tightly bound, quasi-one-dimensional solid: the intense field confines electron orbitals to magnetic-length-scale cylinders aligned with the field, so that atoms bind into linear chains and chains cohere into three-dimensional condensed matter \cite{Chen1974}. These cylindrical atoms behave as elongated electric dipoles aligning into infinite polymeric chains along the field --- a picture that underpins models of pulsar polar-gap accelerators and surface emission \cite{Ruderman1975}. While early estimates relied on semiclassical and Thomas--Fermi-type arguments, quantitatively resolving the exchange and correlation energies of these lattices has demanded progressively more sophisticated quantum-mechanical treatments. We stress that the regime of interest here ($B \gtrsim 10^{12}$ G, far above the atomic reference field $B_0 = 2.35\times10^{9}$ G) is governed by the adiabatic approximation, in which all electrons occupy the lowest Landau level and the electronic structure reduces to an effectively one-dimensional density-functional problem. This is conceptually distinct from current-density functional theory (CDFT) \cite{Vignale1987, Penz2023}, the rigorous framework for the laboratory and molecular regime ($B \lesssim B_0$) in which the paramagnetic current density remains an essential variable; in the neutron-star regime adopted here, a density-only functional of the magnetized electron gas is sufficient.

Computational approaches to this problem have developed along two complementary lines. The first targets condensed surface matter directly: \citet{Flowers1977} produced the first variational (Thomas--Fermi-type) estimates of the cohesive energy of magnetized iron, and \citet{Jones1985} performed the earliest density-functional band-structure calculations of atoms and infinite linear molecules at neutron-star field strengths, already noting the difficulty of capturing electron correlation in this regime. A second line, rooted in atomic and molecular physics, applied Hartree--Fock and configuration-interaction methods to the binding energies of atoms and short molecular chains \cite{Neuhauser1986}. These threads were unified and substantially extended by \citet{medin2006finite, medin2006periodic}, who developed a self-consistent density-functional framework spanning isolated atoms, molecules, and infinite chains, and applied it to condensed surfaces and pulsar polar-cap physics \cite{medin2007condensed, medin2008thesis}. More recently, highly optimized Hartree--Fock--Roothaan codes have been built to generate large atomic databases for the spectral modeling of magnetic white dwarfs and neutron stars \cite{Schimeczek2014}. 

Despite this progress, openly available, extensible, and well-documented community codes for condensed matter in the neutron-star regime remain scarce --- most existing implementations are proprietary, restricted to isolated atoms, or no longer maintained. \codename\ is designed to fill this gap: it recasts the Medin--Lai adiabatic DFT on a real-space grid, providing a transparent, modular, open-source solver for one-dimensional atomic chains in fields of $10^{12}$--$10^{15}$ G.

We benchmark UnReal-B against established calculations
in the literature and demonstrate that it reproduces key structural
and energetic properties with high accuracy. 
By emphasizing
simplicity, reproducibility, and extensibility, UnReal-B provides a
practical framework for future investigations of strongly magnetized
condensed matter and its observational signatures.

\section{Methodology}
The DFT methodology is primarily based on the foundational work proposed by \cite{medin2006periodic,medin2006finite} - which we shall avoid repeating here in details. The formalism is used for 1D atomic chains along the magnetic field lines. At the core of the formalism is the `adiabatic principle' which assumes that under extreme magnetic fields ($B \gtrsim 10^9$ G) the magnetic length $a_0 = \sqrt{\hbar/eB}$ is significantly smaller than the Bohr radius and all the electrons occupy lowest Landau level. In the strong-field regime, the transverse degrees of freedom are described by magnetic orbitals $W_m(\mathbf{r}_\perp)$, which represent the degenerate states within the lowest Landau level manifold. The total three-dimensional wavefunction for an electron in the $m$-th orbital and $\nu$-th band is:
\begin{equation}
\Psi_{m\nu k}(\mathbf{r}) = W_m(\mathbf{r}_\perp) \psi_{m\nu k}(z)
\end{equation}

To exploit the periodic symmetry of the 1D chain, the longitudinal component is expressed in Bloch form:
\begin{equation}
\psi_{m\nu k}(z) = \frac{1}{\sqrt{N}} e^{ikz} u_{m\nu k}(z)
\end{equation}
where $u_{m\nu k}(z)$ is the cell-periodic envelope function. The core of the electronic structure problem relies on solving the effective one-dimensional Kohn-Sham equation for each magnetic orbital $m$ and wavevector $k$:
\begin{equation}
\left[ \hat{T}_k + V_{\text{eff},m}(z) \right] u_{m\nu k}(z) = \epsilon_{m\nu k} u_{m\nu k}(z)
\end{equation}
where $\epsilon_{m\nu k}$ are the corresponding single-particle energy eigenvalues and $\hat{T}_k = \frac{1}{2}(-i\partial_z + k)^2$ is the kinetic energy operator. To solve this numerically, the longitudinal coordinate is mapped to a real-space grid $z_i \in [-a/2, a/2]$ with $N_z$ points, and the kinetic operator is discretized using a finite difference scheme. This representation allows the effective Hamiltonian to be treated as a sparse Hermitian matrix, which is then diagonalized using standard numerical libraries.

The effective potential $V_{\text{eff},m}(z)$ is constructed as the sum of ionic, Hartree, and exchange-correlation contributions, all of which are averaged over the transverse Landau probability density $|W_m(\mathbf{r}_\perp)|^2$:
\begin{equation}
V_{\text{eff},m}(z) = V_{\text{ion},m}(z) + V_{\text{H},m}(z) + V_{\text{xc},m}(z)
\end{equation}

The ionic potential $V_{\text{ion},m}(z)$ represents the interaction between the electron in orbital $m$ and the periodic array of nuclei. It is calculated by explicitly summing the contributions from the central nucleus and $N_Q$ neighboring cells:
\begin{equation}
V_{\text{ion},m}(z) = -Ze^2 \sum_{j=-N_Q}^{N_Q} \int d\mathbf{r}_\perp \frac{|W_m(\mathbf{r}_\perp)|^2}{\sqrt{r_\perp^2 + (z - ja)^2}}
\end{equation}
By restricting the explicit summation to the near-field (typically $N_Q=1$), we maintain high numerical precision near the nuclei where the potential varies most rapidly. The 1D Coulomb divergence of the infinite nuclear chain is naturally managed by the combined near-field electronic potential and the far-field quadrupole correction described in the following paragraphs. 

The electron-electron (Hartree) interaction represents the primary computational bottleneck, and for the $m$-th magnetic orbital it is defined as the sum of near-field and far-field contributions:
\begin{equation}
V_{\text{H},m}(z) = V_{\text{near},m}(z) + V_{\text{far},m}(z)
\end{equation}

The near-field component, $V_{\text{near},m}(z)$, represents the exact electrostatic repulsion from the electronic density within the central unit cell and its $N_Q$ immediate neighbors. It is computed as the sum of 1D convolutions between the longitudinal densities $n_{m'}(z')$ and the transverse-averaged Laguerre kernels $K_{m,m'}$:
\begin{equation}
\begin{split}
V_{\text{near},m}(z) = & e^2 \sum_{m'=0}^{M_{\text{max}}-1}  \sum_{j=-N_Q}^{N_Q}  \\
& \int_{-a/2}^{a/2} dz'  \times K_{m,m'}(|z - z' - ja|) n_{m'}(z')
\end{split}
\end{equation}
In the  work of  \cite{medin2006periodic}, this Hartree potential was evaluated via explicit, point-by-point numerical quadrature. \codename\ advances this by reframing the integration as a global, grid-based convolution, providing two distinct solver options. In the direct matrix-vector multiplication approach, the kernels $K_{m,m'}$ are pre-evaluated as static matrices, reducing the near-field potential evaluation to a sequence of highly optimized matrix-vector multiplications using standard libraries. Alternatively, for high-resolution spatial grids, the code offers an accelerated linear fast Fourier transform convolution. By performing a linear convolution over a tiled spatial domain, this method avoids the unphysical wrap-around artifacts and DC offsets characteristic of standard circular convolutions.

The long-range contribution $V_{\text{far},m}(z)$ from the remaining infinite chain ($|j| > N_Q$) is calculated by expanding the potential of the neutral unit cell into its multipole components. Following \cite{medin2006periodic}, we define the macroscopic quadrupole moment $Q_{zz}$:
\begin{equation}
Q_{zz} = \sum_{m} \int_{-a/2}^{a/2} dz \, n_m(z) \left( 2z^2 - 2(m+1)\rho_0^2 \right)
\end{equation}

The far-field correction is added analytically, using the Hurwitz zeta function to sum the exterior cells:
\begin{equation}
\begin{split}
V_{\text{far},m}(z) &= 3 Q_{zz} \frac{\zeta(5, N_Q+1)}{a^5} \left( 2z^2 - 2(m+1)\rho_0^2 \right) \\ 
&+ 6 Q_{zz} \frac{\zeta(3, N_Q+1)}{a^3}
\end{split}
\end{equation}
As the nuclear point charges sitting at the origin do not contribute to the macroscopic quadrupole moment of the cell ($Q_{\text{ion}} = 0$), applying this correction to the electronic potential alone fully accounts for the charge-neutral infinite periodic environment.

Following the evaluation of the classical electrostatic interactions, the non-classical many-body effects are incorporated within the strong-field Local Density Approximation (LDA). In this framework, the local exchange-correlation potential is formally equivalent to the exchange-correlation chemical potential of a homogeneous electron gas, evaluated at the local 3D density: $V_{\text{xc}}^{\text{3D}}(\mathbf{r}) \equiv \mu_{\text{xc}}(n(\mathbf{r}))$. The precise functional form for $\mu_{\text{xc}}$ implemented in \codename\ follows the comprehensive parameterization derived by  \cite{medin2006periodic}. As their analytical fitting functions—which smoothly interpolate between the low-density and high-density limits of a highly magnetized electron gas—are extensive, we direct the reader to their foundational work for the complete expressions. Because the local density $n(\mathbf{r}) = \sum_{m} |W_m(\mathbf{r}_\perp)|^2 n_m(z)$ couples all occupied magnetic orbitals, the effective 1D potential for the $m$-th orbital cannot be decoupled analytically. Instead, it is obtained by averaging $\mu_{\text{xc}}$ over the corresponding transverse probability density:
\begin{equation}
V_{\text{xc},m}(z) = \int_{0}^{2\pi} \int_{0}^{\infty} |W_m(\rho, \phi)|^2 \mu_{\text{xc}}(n(\rho, z)) \, \rho \, d\rho \, d\phi
\end{equation}
In \codename, this transverse integration is evaluated numerically on a discretized $(\rho, z)$ grid at each step of the self-consistent field iterations. This ensures that the dynamic spatial variations of exchange and correlation are accurately captured as the coupled longitudinal densities evolve toward the ground state.

The total energy per unit cell of the 1D chain is given by:
\begin{equation}
\begin{split}
E_{\text{tot}} =& E_{\text{band}} + E_{\text{ion-ion}} - \\ &  \sum_m \int_{-a/2}^{a/2} dz \, n_m(z) \left[ \frac{1}{2} V_{\text{H},m}(z) + V_{\text{xc},m}(z) \right] + E_{\text{xc}}
\end{split}
\end{equation}
where $E_{\text{band}}$ is the sum of the occupied single-particle energies weighted by their respective fractional occupations $\omega_{m\nu}$:
\begin{equation}
E_{\text{band}} = \sum_{m, \nu} \omega_{m\nu} \epsilon_{m\nu}
\end{equation}
The many-body exchange-correlation energy $E_{\text{xc}}$, which is evaluated by integrating the local energy density $\epsilon_{\text{xc}}$ over the unit cell is:
\begin{equation}
E_{\text{xc}} = \sum_m \int_{-a/2}^{a/2} dz \iint_{\mathbb{R}^2} d\mathbf{r}_\perp |W_m(\mathbf{r}_\perp)|^2 n_m(z) \epsilon_{\text{xc}}(n(\mathbf{r})).
\end{equation}

The structural ground state of the chain is ultimately determined by iteratively solving the 1D Kohn-Sham equations to self-consistency, locating the configuration where $E_{\text{tot}}$ is minimized.

As justified in earlier works, within the lowest Landau level ($n=0$) with complete spin alignment ($\sigma_z = -1/2$), the magnetic field contribution to the relativistic energy spectrum explicitly cancels out. Consequently, the transverse zero-point motion acquires no relativistic mass penalty, and relativistic kinematics are entirely confined to the longitudinal momentum. For the lighter elements and moderate field regimes investigated, the longitudinal Fermi momentum remains deep within the non-relativistic limit ($p_F \ll m_e c$), rendering the above prescribed  DFT treatment, accurate enough description of the electronic structure.
 
\codename\  is implemented in Python, leveraging the NumPy and SciPy ecosystems for linear algebra and integration. The code is modular, allowing users to easily swap exchange-correlation functionals or modify the unit cell geometry. Key routines for potential evaluation are optimized using vectorized operations, making the solver light enough to run on standard desktop hardware while maintaining the precision required for astrophysical research.

\begin{table*}[t]
\centering
\label{tab:data_structure}
\begin{tabular}{llp{7.5cm}l}
\toprule
\textbf{Category} & \textbf{Key / Path} & \textbf{Description} & \textbf{Example / Unit} \\ 
\midrule
\rowcolor[gray]{0.9} \multicolumn{4}{l}{\textit{Input Configuration (YAML)}} \\
\texttt{grid}       & \texttt{Nk, Nz}        & Number of $k$-points and longitudinal grid points. & 40, 501 \\
\texttt{simulation} & \texttt{B\_field}      & Magnetic field strength. & $1.0 \times 10^{15}$ G \\
                    & \texttt{atomic\_Z}     & Nuclear charge ($Z$) and electron count. & 6 \\
                    & \texttt{lattice\_const}& Periodic spacing ($a$). & 0.022 a.u. \\
                    & \texttt{M\_max, nu\_max}& Total magnetic orbitals and longitudinal bands.\footnote{In the config file, these parameters define the total count of states rather than the maximum index.} & 69, 1 \\
                    & \texttt{use\_fft}      & Toggle for the $\mathcal{O}(N \log N)$ Hartree solver. & \texttt{true} \\
\texttt{scf}        & \texttt{mixing, tol}   & Density mixing ($\alpha$) and energy tolerance. & 0.25, $10^{-3}$ \\
\texttt{output}     & \texttt{save\_data}    & Flag to export binary HDF5 results. & \texttt{true} \\
\midrule
\rowcolor[gray]{0.9} \multicolumn{4}{l}{\textit{Output Data Archive (HDF5)}} \\
\texttt{/metadata}  & (Attributes)           & Global parameters, timestamps, and hardware metadata. & --- \\
\texttt{/grid}      & \texttt{z, k}          & Numerical grids for real-space and $k$-space. & Bohr, a.u. \\
\texttt{/density}   & \texttt{total, partial}& Total $n(z)$ and band-resolved $n_m(z)$ densities. & $a_0^{-1}$ \\
\texttt{/eigenvalues}& ---                   & Band energies $\varepsilon_{m,\nu}(k)$; shape ($M_{\text{max}}, \nu_{\text{max}}, N_k$). & eV \\
\texttt{/eigenvectors}& ---                  & Complex longitudinal wavefunctions $f_{mk}(z)$. & Complex128 \\
\texttt{/occupations}& \texttt{sigma, k\_f}  & Fractional occupations and Fermi wave-numbers $k_F^m$. & ---, a.u. \\
\texttt{/potentials} & \texttt{V\_eff, V\_ee}& Effective, Hartree, and Exchange-Correlation potentials. & Hartree \\
\texttt{/energy}     & (Attributes)          & Breakdown of total energy components. & eV \\
\texttt{/convergence}& \texttt{errors, Etot} & Iteration history of density residuals and energy. & ---, eV \\
\bottomrule
\end{tabular}
\caption{Comprehensive data structure of \codename: Input configuration keys and corresponding HDF5 output datasets.}
\label{tab:IO}
\end{table*}

\subsection{Mechanical and thermodynamic properties}
To extract macroscopic and thermal properties from the calculated total energies, the energy-lattice constant relation $E(a)$ is fitted to a third-order polynomial near the minimum:$$ E(a) = c_3 a^3 + c_2 a^2 + c_1 a + c_0 $$This analytical form allows for the determination of the equilibrium lattice constant $a_{eq}$ by solving the condition $dE/da = 0$. The curvature and higher-order derivatives at this minimum define the effective spring constant $k$ and the anharmonicity parameter $\alpha$:$$ k = \frac{d^2E}{da^2}\bigg|_{a_{eq}}, \quad \alpha = \frac{d^3E}{da^3}\bigg|_{a_{eq}} $$The dimensionless Grüneisen parameter $\gamma$, which characterizes the lattice response to thermal expansion and the pressure-volume relationship, is derived from the ratio of these derivatives:$$ \gamma = - \frac{a_{eq}}{2k} \alpha $$Modeling the 1D chain as a series of harmonic oscillators with mass $M$, the characteristic vibrational frequency, within Einstein solid approximation\footnote{Because a full density functional perturbation theory (DFPT) calculation of the phonon density of states is computationally expensive and is beyond the scope of current work, we approximate the characteristic zero-point vibrational frequency using the curvature of the macroscopic equation of state, treating the system within an Einstein-like oscillator framework.}, is $\omega = \sqrt{k/M}$. This frequency defines the longitudinal sound speed $v_s$ and the Debye temperature $\Theta_D$:$$ v_s = a_{eq} \omega, \quad \Theta_D = \frac{\hbar \omega_{max}}{k_B} $$where $\omega_{max}$ is determined by the Brillouin zone boundary. The melting temperature $T_{melt}$ is estimated using a Lindemann-like criterion, assuming the lattice becomes unstable when the root-mean-square displacement of the ions reaches 10\% of the lattice constant:
$$ T_{melt} = \frac{k (0.1 a_{eq})^2}{k_B} $$
The macroscopic mass density $\rho$ is calculated by assuming the transverse area of the unit cell $S$ is governed by the magnetic flux, $S = 2\pi \hat{\rho}^2$, where $\hat{\rho}$ is the magnetic length. Using the atomic mass $M_{amu}$ and the calculated $a_{eq}$:
$$ \rho = \frac{M_{amu}}{S \cdot a_{eq}}. $$

Finally, the cohesive energy $Q_s$, which determines the binding strength of the condensed phase and the ion emission threshold for pulsar magnetospheres, requires careful consideration of lattice vibrations. Previous treatments, such as the semi-analytic models by Medin and Lai, typically approximate the cohesive energy strictly from the electronic potential minimum, evaluating it as $Q_s^0=E_{atom} - E_{eq}$. This approach implicitly neglects the zero-point vibrational energy of the ions, an approximation that is standard in terrestrial solid-state physics where zero-point energies are negligibly small compared to bulk binding energies. However, under extreme magnetic confinement, the 1D atomic chains become exceptionally stiff, leading to high characteristic phonon frequencies and a non-negligible zero-point energy that physically destabilizes the lattice. Therefore, to provide a more rigorous calculation of the binding threshold, our real-space DFT methodology explicitly utilizes the derived spring constant to compute the phonon frequency, allowing us to incorporate the zero-point energy correction into the cohesive energy:$$ Q_s = E_{atom} - (E_{eq} + \frac{1}{2}\hbar\omega). $$
In practice, the atomic energy ($E_{\text{atom}}$) is computed within the same periodic framework using a supercell approach, where a very large lattice constant ($a \rightarrow \infty$) effectively isolates the atom. This also ensures the consistent cancellation of systematic errors.

These derived quantities, computed with quantum vibrational corrections, serve as direct and highly accurate inputs for modeling the neutron star mass-radius relationship, the depth of the solid-liquid transition in the crust, and the feasibility of vacuum gap accelerators in pulsar magnetospheres.

\section{Structure of the code}
The \codename\ code is organized in separate python \texttt{class}es for ease of portability. The main \texttt{class} being the \texttt{SCFDriverPeriodicMultiBand} which governs the self-consistency loop.   
During each iteration, the class builds and diagonalizes the Hamiltonian for every Landau level $m$ to obtain the multi-band spectrum. A single global Fermi energy is then determined via Brent’s algorithm to satisfy the total charge neutrality condition. Once this Fermi level is found, the specific occupations ($\sigma$) and Fermi momenta ($k_F$) for each $(\nu, m)$ band are calculated based on where that global level intersects each band's dispersion.

Numerical experiments in $\codename$ are governed by a central configuration file in YAML format. To ensure full reproducibility, this file decouples the physical model from the numerical execution parameters. The parameters are organized into four primary sections as described in Table \ref{tab:IO}. There is no specific order for the blocks or the parameters keys. A summary of a typical configuration for a condensed Carbon chain ($Z=6$) in a magnetar-strength field is provided in Table~\ref{tab:IO}.

\textbf{The \texttt{simulation} Section}\ This block defines the fundamental physics of the system. The magnetic field strength (\texttt{B\_field}) is specified in Gauss, which determines the magnetic length and Landau level spacing. The nuclear charge (\texttt{atomic\_Z}) and \texttt{num\_electrons} define the atomic species. The \texttt{lattice\_constant} set the periodicity of the 1D chain. The model's complexity is further controlled by \texttt{M\_max} (the number of magnetic orbitals) and \texttt{nu\_max} (the number of longitudinal bands).

\textbf{The \texttt{grid} Section}\ The discretization of the system is defined by \texttt{Nz}, the number of spatial points along the $z$-axis within the unit cell, and \texttt{Nk}, the number of $k$-points used to sample the 1D Brillouin zone.

\textbf{The \texttt{scf} Section}\ The self-consistent field iterations are regulated by \texttt{mixing}, which linearly blends the density from previous iterations to stabilize convergence. The simulation terminates when the change in energy falls below the \texttt{tol} (tolerance) or when \texttt{max\_iter} is reached.

\textbf{The \texttt{output} Section}
The final block allows users to store the data for further processing using the key \texttt{save\_data}. The output saved as a compressed HDF5 (Hierarchical Data Format) archive. Unlike flat-file formats, the HDF5 structure allows for the storage of high-dimensional datasets—such as the complex-valued longitudinal wavefunctions—alongside scalar metadata and convergence diagnostics. The output file is organized into a hierarchical tree, as summarized in Table~\ref{tab:IO}. This structure allows for "lazy loading," where specific datasets (e.g., the total density) can be accessed for plotting without loading the memory-intensive wavefunctions.

All multidimensional arrays, such as the eigenvectors of shape $(M_{\text{max}}, N_k, N_z)$, are stored using GZIP compression (level 4) to significantly reduce the on-disk footprint without loss of numerical precision.

An example configuration file (to be named as \texttt{conf.yaml}) is shown in Appendix \ref{sec:config}.
A separate post-processing script is also provided to demonstrate the usage of the stored data and generation of meaningful plots like band-structure.

\begin{figure}
    \centering
    \includegraphics[width=0.45\linewidth]{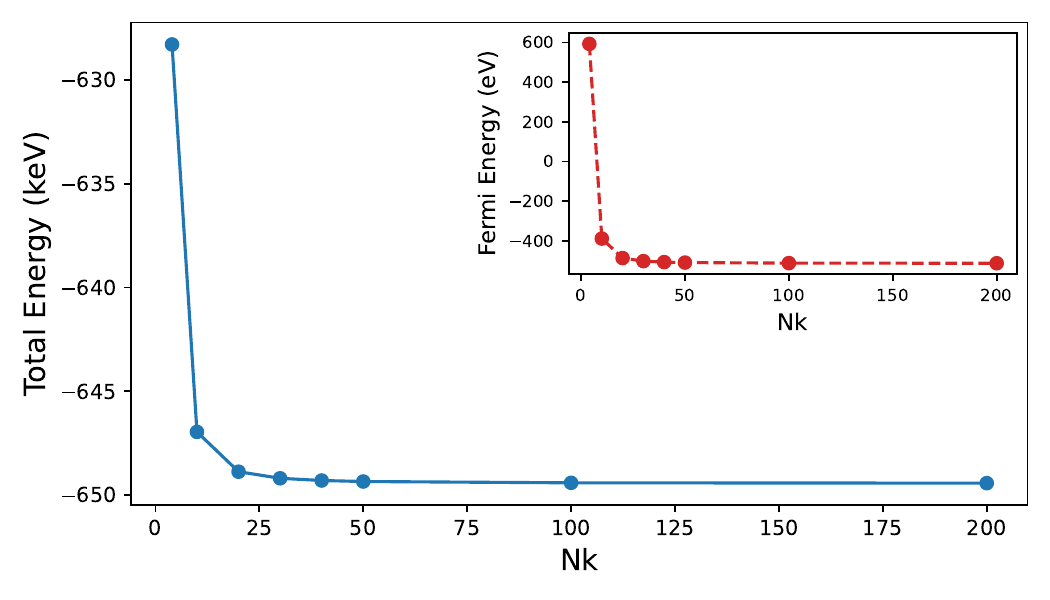}
    \includegraphics[width=0.45\linewidth]{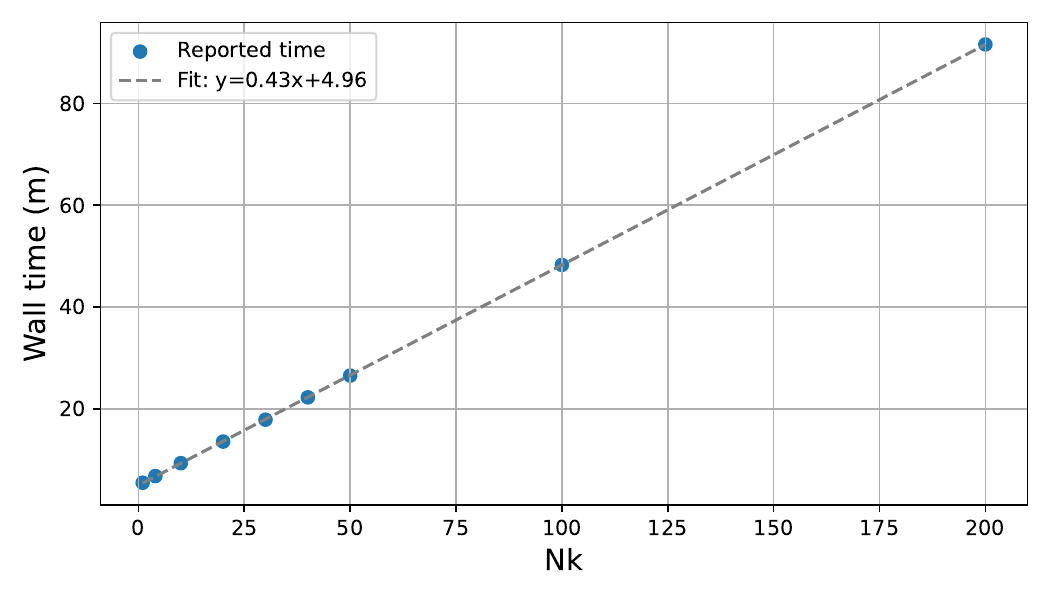}\\
       \includegraphics[width=0.45\linewidth]{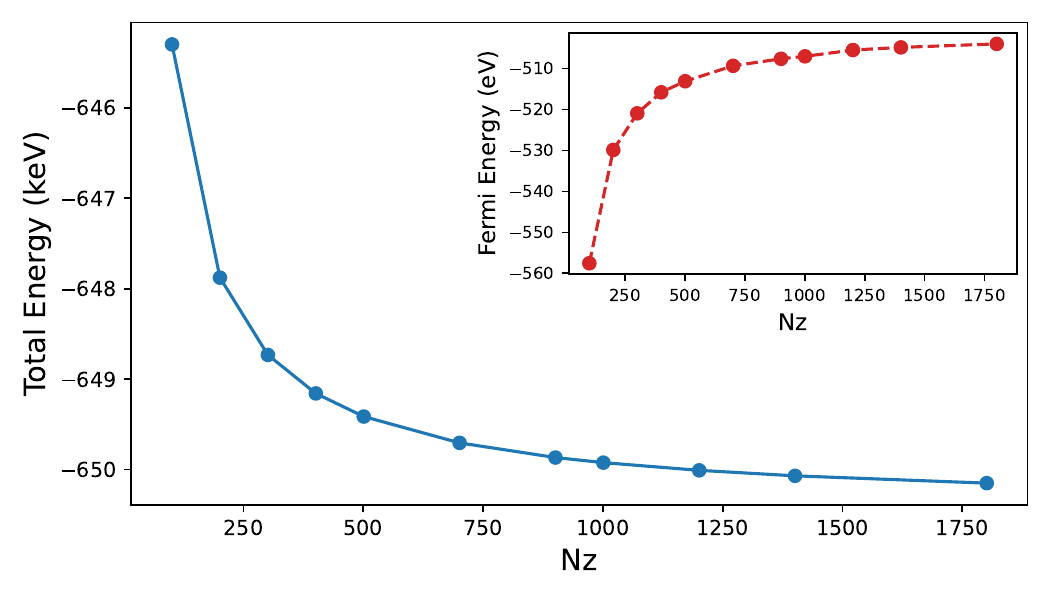}
    \includegraphics[width=0.45\linewidth]{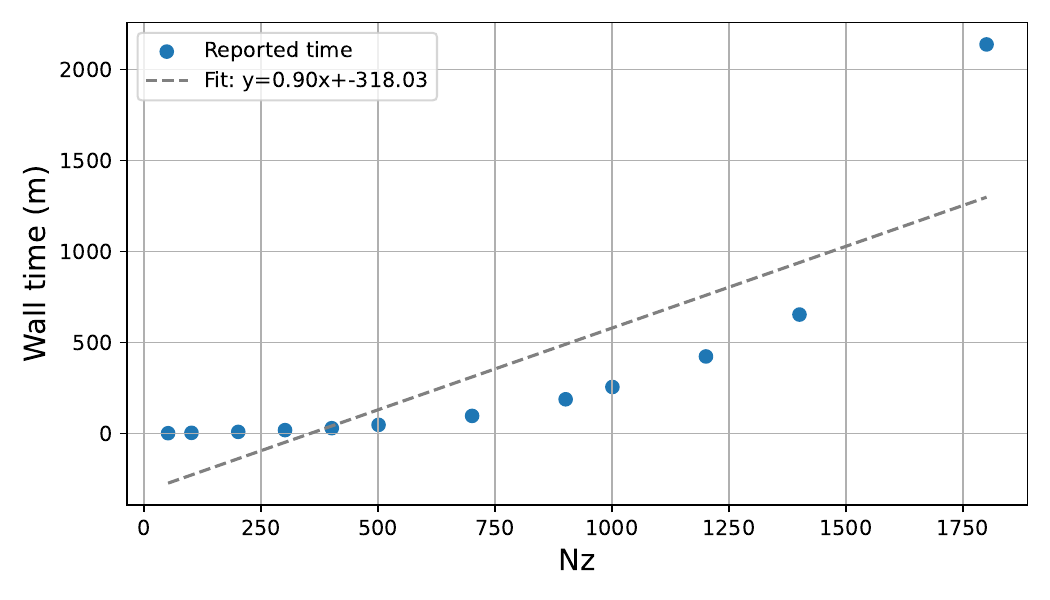}
    \caption{a) Convergence against k-points. b) Computational scaling c) Convergence against z-grid. }
    \label{fig:k_convergence}
\end{figure}

\section{Results and Discussion}
In this section, we discuss a typical DFT workflow with \codename, comparing the obtained results against available benchmarks throughout, with more comprehensive benchmarks presented toward the end of the section. In the discussion below the magnetic field strength $B$ is expressed in units of $10^{12}$ G, denoted by the dimensionless parameter $B_{12} = B / 10^{12}$ G.

\subsection{Convergence}
Before proceeding with any DFT  simulation, it is imperative to conduct a thorough convergence tests. Here, two critical discretization grids govern the numerical accuracy: the real-space $z$-grid, which governs the discretization along $z$ axis, and the reciprocal-space $k$-grid, which samples the momentum space. Both grids are uniformly discretized, with the $z$-grid divided into $Nz$ points and the $k$-grid into $Nk$ points. Convergence must be systematically verified for both grids to avoid artifacts or inaccuracies stemming from insufficient resolution. As demonstrated in the case of an iron chain subjected to a magnetic field of $B_{12}$ = 500, the convergence behavior is illustrated in Figure \ref{fig:k_convergence}(a). Notably, both the total energy and the Fermi energy exhibit rapid convergence with respect to the $k$-grid requiring as few as $Nk = 30$ points to achieve a relative error of just 0.036\% compared to finest grid of $Nk = 200$. We have checked that this pattern is robust under different field strengths and for all the  chains tested. This rapid convergence even under such a high magnetic field is of convenience for computationally efficient simulations without sacrificing accuracy. The Fig (\ref{fig:k_convergence})(b) also shows the time required to complete the simulation. As can be seen The runtime exhibits a clear linear dependence, well described by $t \approx 0.43 Nk + 4.96 $min. As the present implementation is serial, this behavior reflects a direct loop over k-points, with each k-point contributing approximately a constant computational cost. The slope corresponds to the average time per k-point, while the finite intercept represents a fixed initialization overhead. The strictly linear scaling indicates the absence of additional algorithmic bottlenecks, while also implying that the total runtime increases proportionally with  $Nk$. Such a linear behavior holds true for all the cases investigated.

In contrast to the rapid $k$-grid convergence, the real-space $z$-grid requires significantly higher resolution to resolve the sharp electronic features induced by strong magnetic confinement and is indeed case sensitive. As shown in our tests, for iron chain under $B_{12}=500$,  the absolute total energy decreases monotonically as $N_z$ is increased from 51 to 1801. Due to the intense localization of orbitals along the field lines and the resulting steep density gradients near the nuclei, the absolute energy exhibits a systematic shift of approximately 10 keV across the tested range. The Fermi energy ($E_F$) follows a similar monotonic trend, though it proves more robust than the total energy, indicating that the relative spacing of electronic levels is less sensitive to the grid than the absolute depth of the potential well.While such large absolute shifts are a known characteristic of core-electron representation in real-space grids, the relative precision stabilizes effectively at higher resolutions; we observed a relative energy error of only 0.012\% at $N_z = 1401$ compared to our finest reference. However, this precision comes at a significant computational cost. Unlike the reciprocal-space sampling, which exhibits a strictly linear runtime dependence on $N_k$ due to the independent nature of the $k$-point loops, the runtime scales super-linearly with $N_z$. This behavior reflects the increasing computational complexity of the real-space operations—such as solving the Poisson equation and performing the self-consistent field (SCF) iterations—on a denser coordinate grid. Because the physical quantities of interest, specifically the cohesive energy and the equation of state (EOS), are derived from energy differences, these systematic core-level shifts largely cancel out. Consequently, we identified $N_z \approx 900$ as the optimal threshold for this case, providing the necessary precision to resolve the vibrational and thermal properties of the condensed phase while remaining within a regime of manageable computational scaling. This choice changes case to case basis. 

\subsection{Search for the ground state}

\begin{figure}
    \centering
    \includegraphics[width=0.75\linewidth]{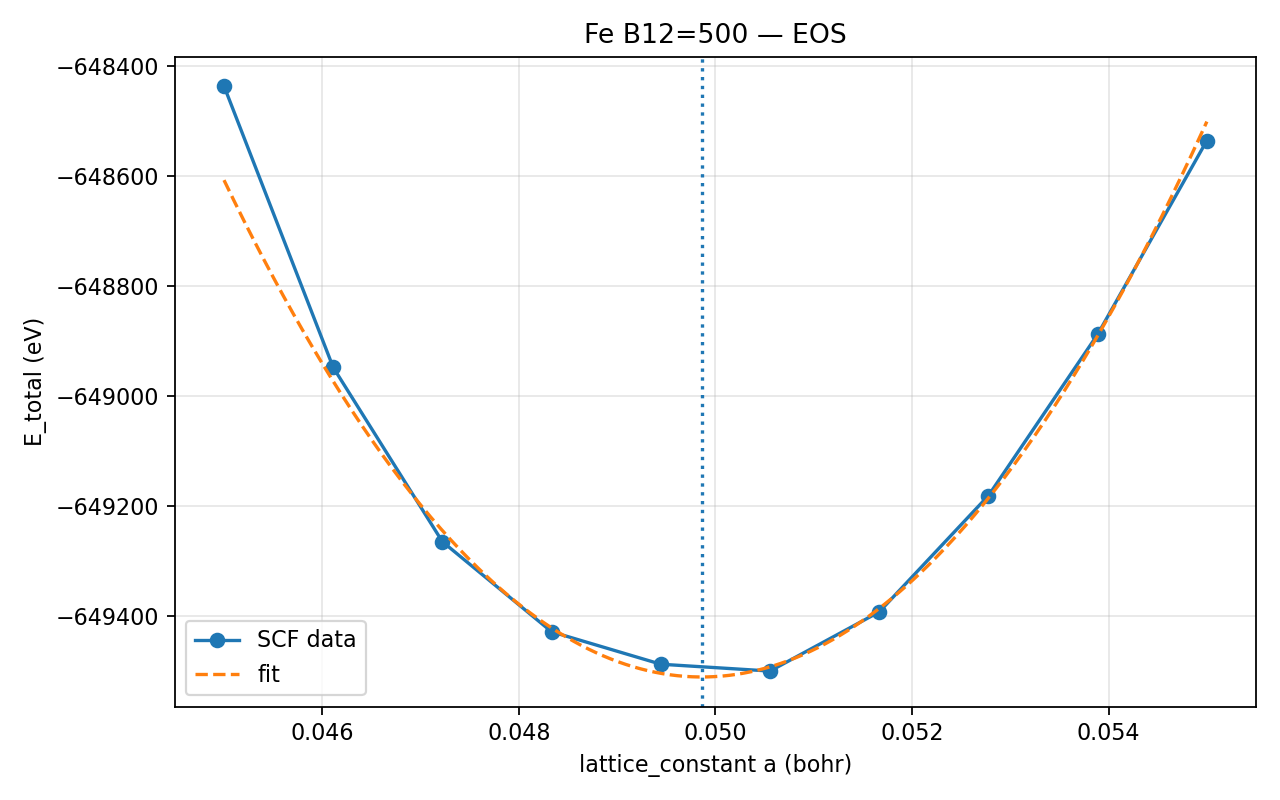}
    \caption{Example of EOS data for Fe, B$_{12}$$=500 $. Points and solid line represent the data while the dotted line is for the quadratic fit.}
    \label{fig:fe_eos}
\end{figure}

\begin{figure}
    \centering
    \includegraphics[width=0.95\linewidth]{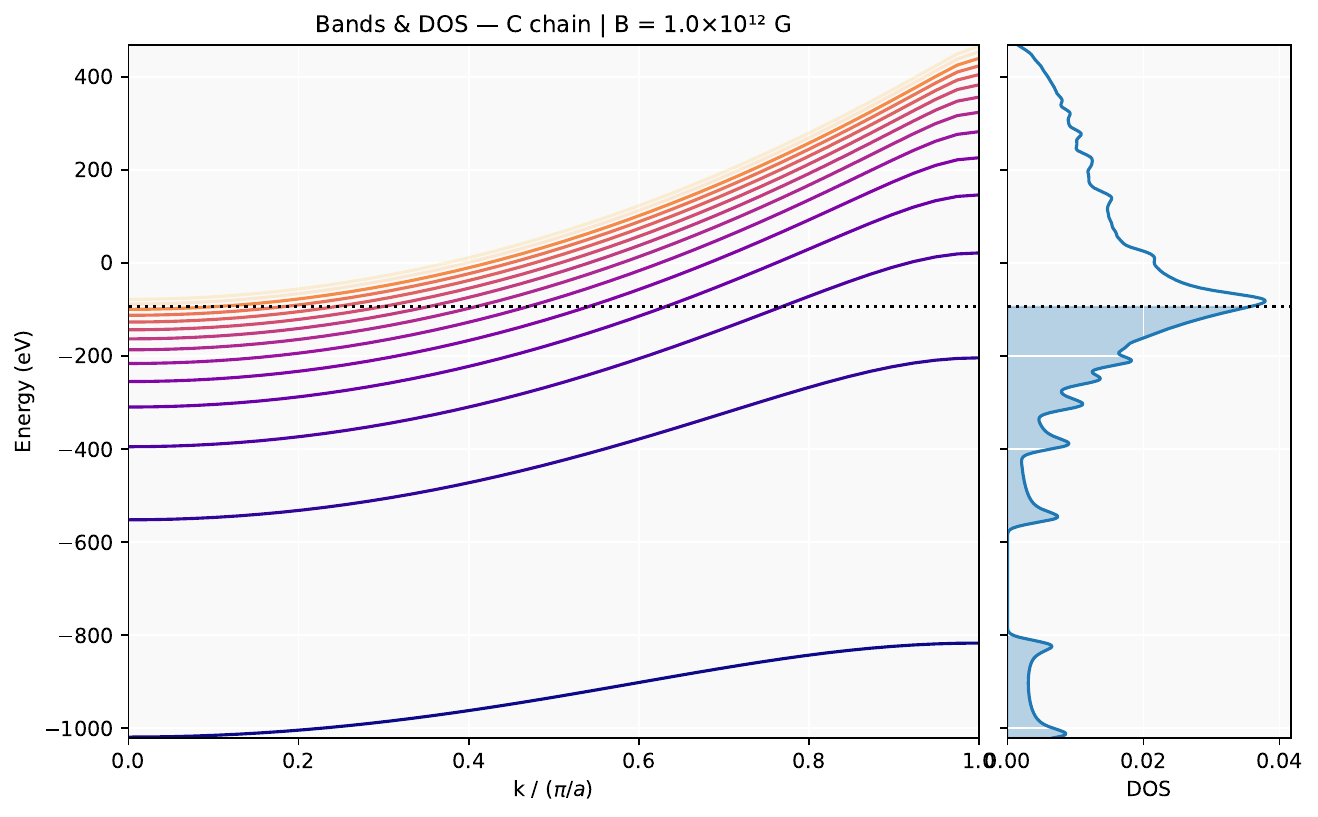}
    \caption{Band Structure of carbon chain. All bands corresponds to $\nu=0$ state. The colors indicate the $m$ values, with red being larger value of $m$. The fermi level is set at 0 eV.}
    \label{fig:c_bands}
\end{figure}

\begin{figure}
    \centering
    \includegraphics[width=0.95\linewidth]{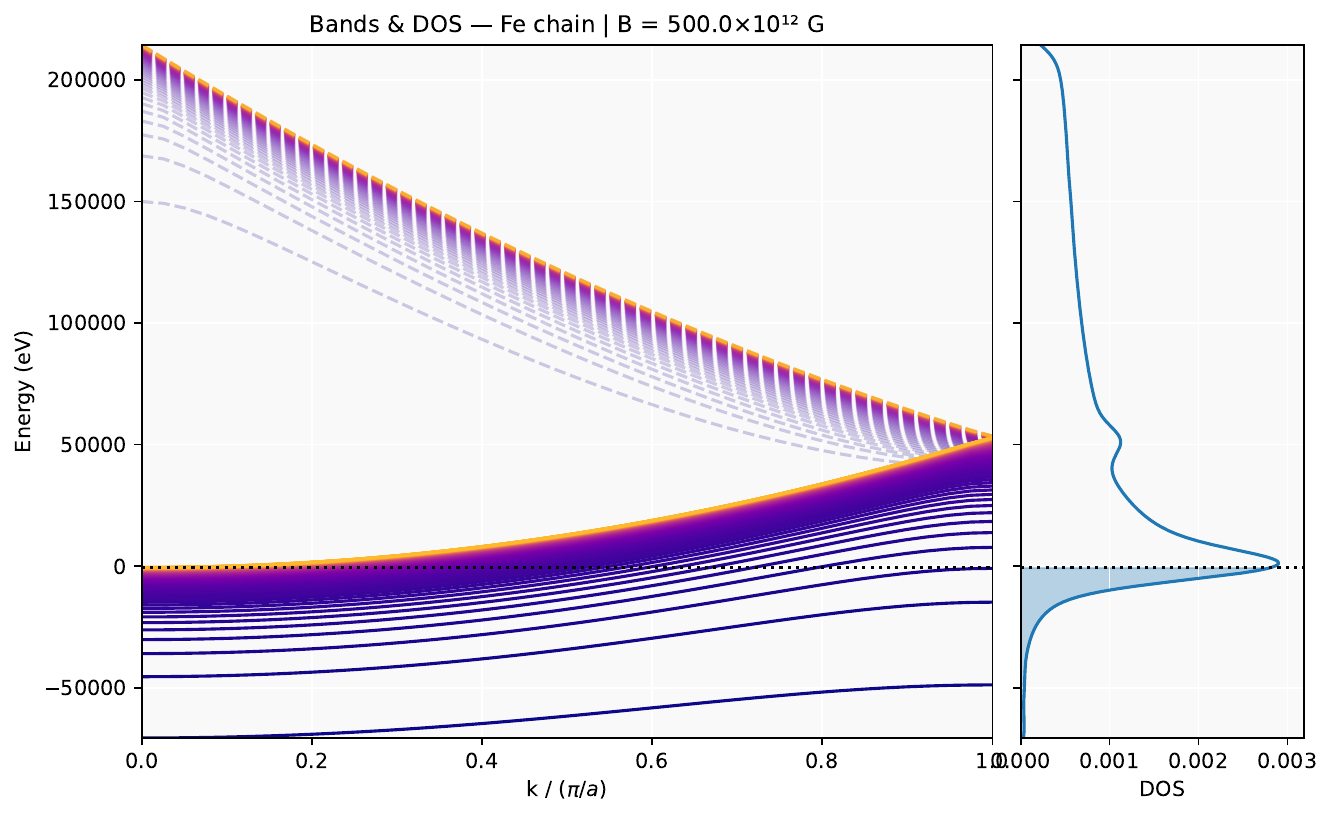}
    \caption{ Band Structure of iron chain under $B_{12}=500$.  The solid lines indicate the $\nu=0$ states and the dashed lines are for $\nu=1$ states. The colors indicate the $m$ values, with yellow being largest value of $m$. The fermi level is set at 0 eV and the number of bands with $\nu=0$ are restricted below fermi level. }
    \label{fig:fe_bands}
\end{figure}

The ground state of the system is determined by various parameters, including the occupied $(m,\nu)$ pairs. While the quantum number $\nu$ typically takes small values (0 or 1), the quantum number $m$ can assume significantly larger values depending on the system. Given sufficiently large values of $(m,\nu)$, \codename\ is capable of identifying filled, partially filled, and unoccupied states. However, we note that computational costs may increase substantially as $m$ grows. 

Beyond the electronic quantum numbers, the macroscopic periodicity of the 1D chain—represented by the lattice constant $a$—must also be determined self-consistently. Because the exact spacing is fundamentally coupled to both the magnetic field strength and the atomic species, we treat $a$ as a variational parameter. We locate the true ground state by systematically scanning a range of trial lattice constants and computing the fully converged total energy, $E_{\text{tot}}(a)$, for each configuration. While this iterative structural relaxation is computationally demanding, it naturally yields the longitudinal Equation of State (EoS) for the magnetic chain. The EoS encodes critical macroscopic physics; for instance, the curvature of the energy profile near the minimum allows us to extract fundamental mechanical and thermodynamic properties, such as the longitudinal bulk modulus and cohesive energy, which we detail in subsequent sections.As a structural benchmark, Figure \ref{fig:fe_eos} illustrates the computed EoS curve for a condensed Iron (Fe) chain subjected to a magnetic field of $B_{12} = 500$. By applying a harmonic (quadratic) fit to the data points clustered around the energy well minimum, we rigorously isolate the equilibrium spacing. This yields a value of $a_{\text{eq}} = 0.0497$ bohr, which is in excellent quantitative agreement with the baseline value of $a_{\text{eq}} = 0.05$ bohr reported by  \cite{medin2006periodic}. This precise match at the macroscopic level provides further validation of both our fine-grid interaction kernels and our comprehensive total energy functional. We have confirmed this level of numerical consistency across the multiple systems investigated in this study the equilibrium lattice constants obtained via \codename\ remain in excellent agreement with previously reported values.

\subsection{Electronic Structure}

To demonstrate the high-resolution capabilities of $\codename$ across vastly different physical regimes, we show, for examples, the longitudinal band structures and corresponding Density of States (DOS) for two systems: a light Carbon chain ($Z=6$) at a baseline magnetar field of $B_{12} = 1$ , and a heavy Iron chain ($Z=26$) at an extreme field of $B_{12} = 500$. 
The electronic spectrum for the condensed Carbon chain ($B = 1.0 \times 10^{12}$ G) is characterized by a set of well-separated, distinct longitudinal subbands spanning from approximately $-1000$ eV to the continuum. As shown in Figure \ref{fig:c_bands}, the lowest-lying core orbitals exhibit small dispersion, indicating stronger localized binding to the nuclei.  The lowest band  can directly be compared to the Figure 3.3 (shown as $\epsilon_{00}$) from the thesis \cite{medin2008thesis}.  On the other hand, near the Fermi level (indicated by 0 eV in the figure), the bands exhibit significant $k$-dispersion, representing valence electrons delocalized along the periodic chain. The DOS panel displays characteristic 1D van Hove singularities at the Brillouin zone boundaries ($k = 0$ and $k = \pi/a$), confirming that the real-space solver accurately captures the sharp density transitions inherent in one-dimensional magnetized matter. The only states that are occupied at $\Gamma$ point are shown the in the figure. 

The results for the Iron chain subjected to $B = 5 \times 10^{14}$ G present a more complex electronic landscape (Fig (\ref{fig:fe_bands}). Due to the intense magnetic confinement and high nuclear charge, the deep core states reside below $-50,000$ eV. In all 105 bands are partially or completely occupied. he solid curves correspond to the ($\nu=0$) Landau manifold for different magnetic quantum numbers (m), while the dashed curves denote the ($\nu=1$) states. 

A notable feature of the calculated spectrum is the apparent energy gap at the $\Gamma$ point ($k=0$). At first glance, this may suggest semiconducting behaviour; however, the corresponding density of states does not exhibit a true band gap. This seemingly counter-intuitive behaviour originates from the fact that the lowest-energy states of the $\nu=0$ manifold are not centered at the Brillouin-zone center. Instead, the occupied bands attain their minima closer to the Brillouin-zone boundary, causing the electronic filling to begin away from $k=0$. Consequently, while a direct gap exists locally at the $\Gamma$-point, states at similar energies remain available at other crystal momenta, thereby preserving a finite density of states at the Fermi level.

\begin{figure}
    \centering
    \includegraphics[width=0.95\linewidth]{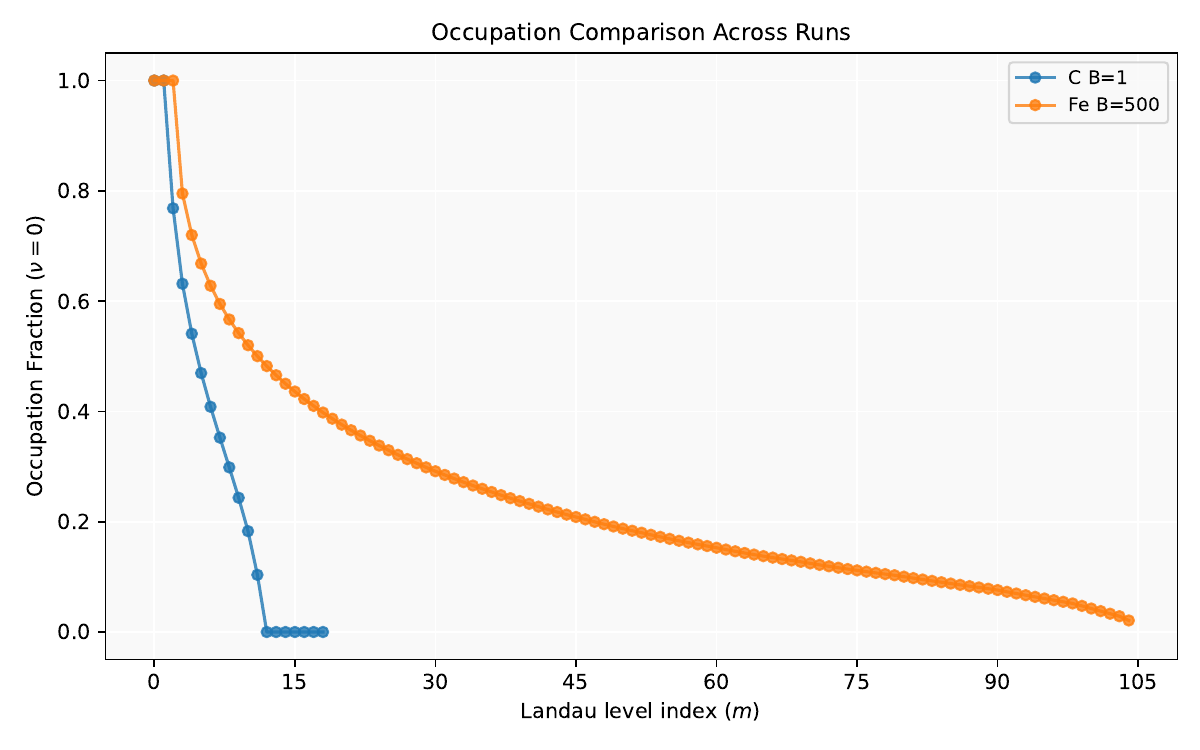}
    \caption{Comparison of band occupancies for carbon chain (blue) under $B_{12}=1$ and iron chain (orange) under $B_{12}=500$. The extreme low values of occupations highlights the high degeneracies, typically for higher values of $B$.}
    \label{fig:occupation}
\end{figure}

The Density of States for the Fe chain underscores the extreme degeneracy of the system, with a massive density peak pinned at the Fermi level. The ability of \codename\  to resolve this dense manifold of states while maintaining numerical stability based linear FFT convolution approach. This degeneracy can further be seen via band occupations.  Fig (\ref{fig:occupation}) compares the occupation fractions of the $\nu=0$ Landau manifold (occupied) as a function of magnetic quantum number $m$ for carbon at $B_{12}=1$ and iron at $B_{12}=500$. While the carbon chain exhibits a relatively compact occupation profile that terminates within a narrow range of low-$m$ states, the Fe chain displays a remarkably extended distribution with finite occupation persisting past $m \sim 100$. The occupations decrease smoothly with increasing $m$, indicating that the electronic structure is distributed over a broad hierarchy of magnetic orbitals rather than being confined to a few tightly bound channels.

This behaviour reflects the strongly anisotropic nature of matter in ultra-intense magnetic fields, where transverse motion is quantized into Landau orbitals while longitudinal motion remains dispersive. 
The smooth occupation profile further supports the metallic nature of the chain inferred from the band structure calculations. 
Although local direct gaps appear at specific crystal momenta, multiple partially occupied $m$-resolved bands remain available near the Fermi level, thereby maintaining a finite density of states. 
The progressively smaller occupation fractions at large $m$ also make the numerical determination of the Fermi energy increasingly delicate, since accurate bracketing of requires resolving a large number of weakly occupied bands with comparable energies.

\begin{figure}
    \centering
    \includegraphics[width=1\linewidth]{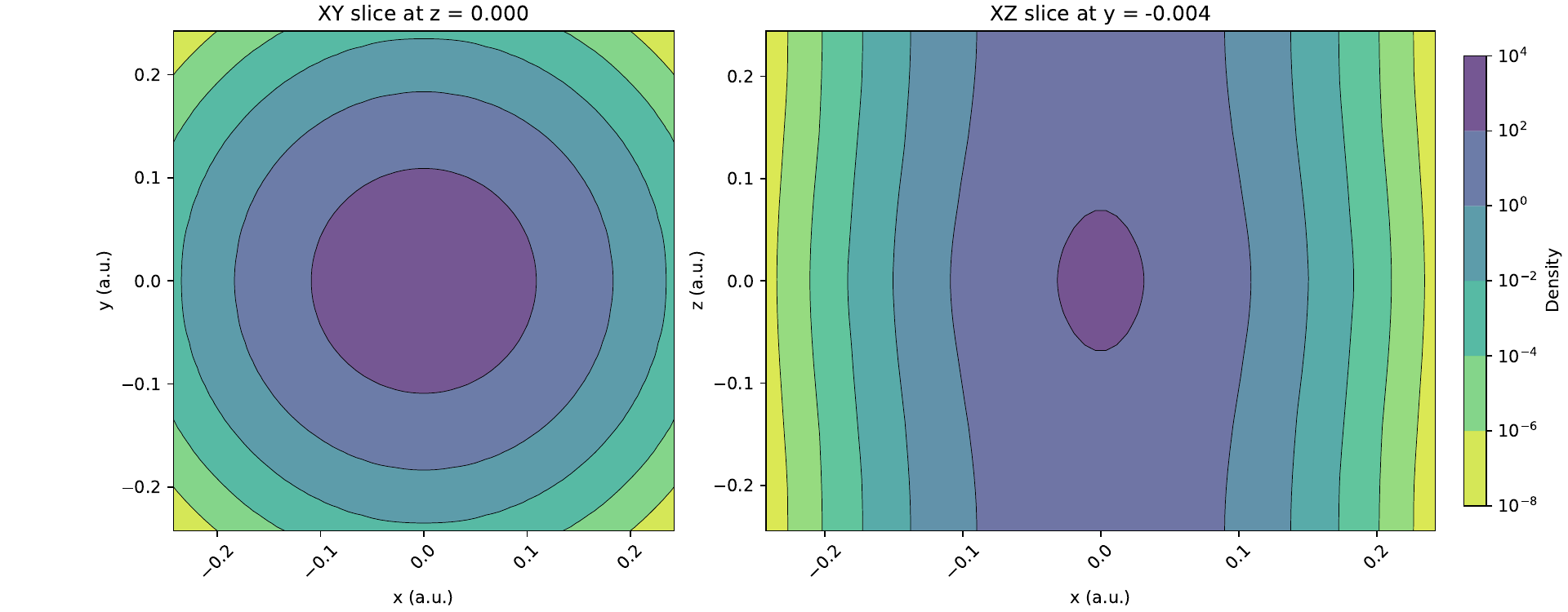}
    \includegraphics[width=1\linewidth]{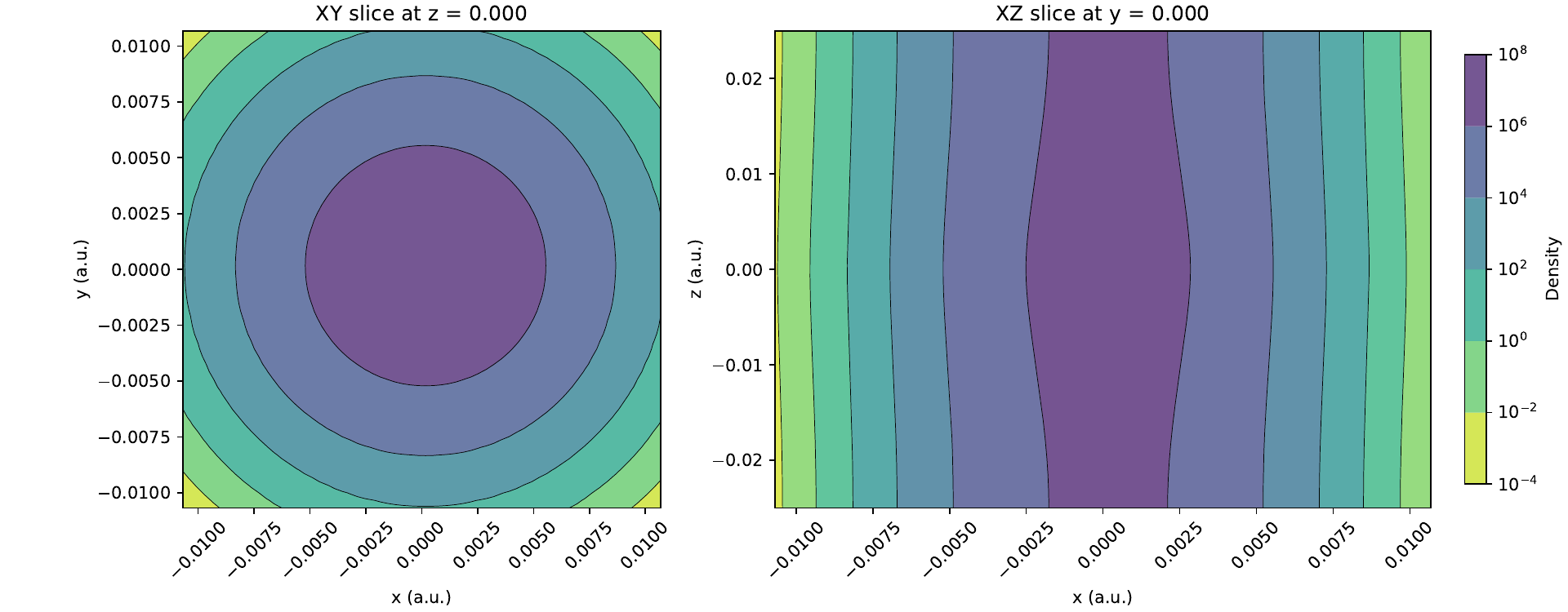}
    \caption{C B$_{12}$=1. Fe B$_{12}=500$. density}
    \label{fig:densities}
\end{figure}

With complete electronic structure in hand we can visualize the charge density. As mentioned before, the \codename\ dumps all the data in hd5 format which can be parsed to plot the 3D densities. As an example we demonstrate the densities for two representative cases. Figure \ref{fig:densities} compares the real-space electronic density slices for carbon at $B_{12}=1$ and iron at $B_{12}=500$. The left panels show transverse $xy$ density slices, while the right panels display longitudinal $xz$ sections through the density maximum. The comparison reveals the dramatic dimensional reduction induced by ultra-strong magnetic fields. In the carbon system, the electronic density retains a relatively broad transverse extent and some anisotropy. In contrast, the Fe chain at $B_{12}=500$ exhibits extreme transverse magnetic compression, with the electronic density confined to a very narrow cylindrical region around the magnetic axis while remaining comparatively extended along the field direction.

This behaviour reflects the dominance of Landau quantization in the ultra-strong-field regime, where the transverse electronic motion is frozen into tightly confined magnetic orbitals. The resulting electronic structure becomes highly anisotropic. The elongated longitudinal density distribution is consistent with the strongly dispersive $\nu=0$ bands observed in the electronic band structure, while the extreme transverse localization illustrates the magnetic confinement responsible for chain formation and anisotropic bonding under neutron-star conditions.

\begin{table*}[]
\centering
\scriptsize
\setlength{\tabcolsep}{3pt}
\begin{tabular}{l c c c c c c}
\hline
El & $B_{12}$ 
& $E_\infty$ (DP / FFT / Ref) 
& $\Delta E_\infty$ (\%) (DP / FFT)
& $\epsilon_F$ (DP / FFT / Ref) 
& $\Delta \epsilon_F$ (\%) (DP / FFT)
& $n_m$ (DP / FFT / Ref) \\
\hline
H & 1 
& -220.47 / -220.52 / -221 
& 0.24 / 0.22 
& -87.05 / -87.13 / -85 
& -2.41 / -2.51 
& 6 / 6 / 6 \\

H & 10 
& -526.50 / -526.63 / -529.2 
& 0.51 / 0.49 
& -171.25 / -171.44 / -165 
& -3.79 / -3.90 
& 10 / 10 / 10 \\

H & 100 
& -1239.56 / -1239.90 / -1253 
& 1.07 / 1.05 
& -323.87 / -324.37 / -311 
& -4.14 / -4.30 
& 16 / 16 / 16 \\

H & 1000 
& -2886.13 / -2886.97 / -2962 
& 2.56 / 2.53 
& -563.24 / -564.44 / -571 
& 1.36 / 1.15 
& 29 / 29 / 26 \\

\hline

He & 1 
& -661.30 / -661.48 / -662.4 
& 0.17 / 0.14 
& -87.38 / -87.49 / -85 
& -2.79 / -2.93 
& 9 / 9 / 9 \\

He & 10 
& -1602.61 / -1603.07 / -1608 
& 0.33 / 0.31 
& -173.23 / -173.54 / -167 
& -3.73 / -3.92 
& 14 / 14 / 14 \\

He & 100 
& -3849.76 / -3850.91 / -3874 
& 0.63 / 0.60 
& -315.28 / -316.09 / -310 
& -1.70 / -1.97 
& 23 / 23 / 23 \\

He & 1000 
& -9209.49 / -9212.37 / -9329 
& 1.28 / 1.25 
& -555.86 / -557.82 / -568 
& 2.14 / 1.79 
& 40 / 40 / 39 \\

\hline

C & 1 
& -4358.73 / -4360.01 / -4367 
& 0.19 / 0.16 
& -93.54 / -93.87 / -92.8 
& -0.79 / -1.15 
& 12 / 12 / 12 \\

C & 10 
& -10287.20 / -10290.48 / -10315 
& 0.27 / 0.24 
& -174.61 / -175.39 / -173 
& -0.93 / -1.38 
& 23 / 23 / 23 \\

C & 100 
& -24950.76 / -24958.86 / -25040 
& 0.36 / 0.32 
& -296.47 / -298.26 / -306 
& 3.12 / 2.53 
& 41 / 41 / 41 \\

C & 1000 
& -60932.93 / -60953.09 / -61320 
& 0.63 / 0.60 
& -480.67 / -484.95 / -539 
& 10.82 / 10.03 
& 70 / 70 / 69 \\
\hline
Fe & 5 
& -107166.85 / -107194.31 / -107230 
& 0.06 / 0.11 
& -158.68 / -160.23 / -161 
& 1.44 / 0.48
& 38 / 38 / 38 \\

Fe & 10 
& -142054.96 / -142093.01 / -142150 
& 0.07 / .04 
& -187.98 / -190.03 / -194 
& 3.11 / 2.04 
& 43 / 43 / 44 \\

Fe & 100 
& -354592.98 / -354704.14 / -355800 
& 0.34 / 0.31 
& -349.33 / -355.17 / -384 
& 9.03 / 7.51 
& 69 / 69 / 69 \\

Fe & 500 
& -648993.70 / -648721.02 / -651900 
& 0.45 / 0.49 
& -496.26 / -468.39 / -583 
& 14.88 / 19.66 
& 106 / 106 / 105 \\

Fe & 1000 
& --- / -838006.69 / -842800 
& --- / 0.57 
& --- / -421.43 / -635 
& --- / 33.63 
& --- / 132 / 130 \\
\hline
\end{tabular}
\caption{Comparison of the total energy per unit cell $E_\infty$, Fermi energy $\epsilon_F$, and number of occupied magnetic orbitals $n_m$ for infinite atomic chains of various elements (El) in different magnetic fields ($B_{12}$. Results obtained using the direct-product (DP) and FFT-based solvers are compared with the reference values reported by  \cite{medin2008thesis}. Percentage deviations are computed relative to the reference data. While both approaches reproduce the reference energies with excellent overall agreement, larger deviations in $\epsilon_F$ appear in the highest-field Fe cases due to the increasing density of weakly occupied and nearly degenerate states near the Fermi level (see text). The `$-$' indicates the specific run was not carried out due to unrealistic time or computational requirements. To maintain the consistent comparison, we have used the lattice constant as prescribed by \cite{medin2006periodic,medin2008thesis}}.
\label{tab:benchmarks}
\end{table*}

\subsection{Benchmarks}
Finally we turn our attention to overall benchmarking of the code. Table \ref{tab:benchmarks} summarizes the calculated properties of infinite atomic chains for H, He, C, and Fe over a broad range of magnetic field strengths extending from $B_{12}=1$ to $B_{12}=1000$. The results obtained using the direct-product (DP) and FFT-based implementations are compared against the reference calculations of Medin and Lai. Overall, both numerical approaches reproduce the reference energies and occupation characteristics with very good agreement across all systems considered, demonstrating the robustness of the present real-space framework for strongly magnetized condensed matter.

The total energy per unit cell $E_\infty$ shows consistently small deviations from the reference values, typically below $1\%$ and remaining within a few tenths of a percent for most cases. The agreement is particularly strong for low- and intermediate-field systems, including H, He, and C chains, where both DP and FFT approaches yield nearly identical results. Even in the most demanding Fe calculations, involving total electronic energies approaching $10^6$~eV in magnitude, the deviations remain comparatively small. This indicates that the present implementation accurately captures the strongly anisotropic electronic structure and the substantial magnetic compression characteristic of neutron-star surface matter.

The number of occupied magnetic orbitals $n_m$ also agrees remarkably well with the reference data. In most cases the occupation structure is reproduced exactly, while the few discrepancies that appear at the highest magnetic fields correspond only to differences of one or two magnetic channels. This consistency is particularly important because the occupation distribution directly determines the reconstructed density, band filling, and Fermi-level position in the quasi-one-dimensional electronic structure.

The Fermi energy $\epsilon_F$ exhibits somewhat larger deviations, especially for high-$Z$ systems at extreme magnetic fields. As mentioned before, this is a reflection the intrinsic numerical sensitivity of the Fermi level in ultra-strong-field condensed matter. In these systems, the occupied electronic manifold spans tens to hundreds of kiloelectronvolts, while the states near $\epsilon_F$ become increasingly dense and weakly occupied. As a result, small variations in occupation bracketing, $k$-point sampling, or band interpolation can shift the inferred Fermi level by several tens  of electronvolts without substantially altering the underlying electronic structure. This effect is particularly evident for Fe at $B_{12}=500$ and $1000$, where multiple nearly degenerate bands accumulate close to the Fermi surface. Ideally, a more physically meaningful comparison metric for the Fermi level in such systems would involve quantities such as the relative position of $\epsilon_F$ within the occupied manifold, integrated DOS occupation, or the topology of the partially occupied bands rather than the absolute energy difference alone. However, such a detailed data is not available in the published literature. 

Despite these larger absolute differences in $\epsilon_F$, the overall electronic structure remains physically consistent. The DP and FFT approaches continue to reproduce similar occupation patterns and total energies, indicating that the discrepancies primarily arise from the delicate placement of the Fermi level within a dense manifold of partially occupied states rather than from major differences in the band structure itself. The increasing deviations at extreme magnetic fields therefore highlight the growing numerical complexity of strongly magnetized heavy-element chains rather than a breakdown of the underlying physical description.

\begin{table*}
\centering
\small
\setlength{\tabcolsep}{4pt}

\begin{tabular}{l r r r r r r r}
\hline
El & $B_{12}$ & $a_{eq}$ & Ref. $a$ & $Q_s^0$ & Ref. $Q_s^0$ & $Q_s$ & $\rho$ \\
 & ($10^{12}$G) & (Bohr) & (Bohr) & (eV) & (eV) & (eV) & (g cm$^{-3}$) \\
\hline

H  &    1.0 & 0.2299 & 0.2300 &    48.68 &    59.60 &    45.20 & 3.327$\times10^{3}$ \\
H  &   10.0 & 0.0919 & 0.0910 &   190.87 &   219.70 &   177.28 & 8.319$\times10^{4}$ \\
H  &  100.0 & 0.0373 & 0.0370 &   634.66 &   712.70 &   582.04 & 2.050$\times10^{6}$ \\
H  & 1000.0 & 0.0155 & 0.0145 &  1752.20 &  2092.40 &  1559.21 & 4.948$\times10^{7}$ \\

He &    1.0 & 0.2803 & 0.2800 &    57.84 &    58.90 &    55.46 & 1.083$\times10^{4}$ \\
He &   10.0 & 0.1093 & 0.1090 &   351.67 &   356.00 &   341.44 & 2.779$\times10^{5}$ \\
He &  100.0 & 0.0435 & 0.0430 &  1492.11 &  1489.00 &  1450.96 & 6.977$\times10^{6}$ \\
He & 1000.0 & 0.0176 & 0.0175 &  5138.17 &  5107.00 &  4978.20 & 1.723$\times10^{8}$ \\

C  &    1.0 & 0.4923 & 0.4900 &    18.30 &    26.00 &    17.16 & 1.851$\times10^{4}$ \\
C  &   10.0 & 0.1549 & 0.1540 &   224.13 &   240.00 &   216.55 & 5.883$\times10^{5}$ \\
C  &  100.0 & 0.0564 & 0.0560 &  3676.59 &  3680.00 &  3631.01 & 1.616$\times10^{7}$ \\
C  & 1000.0 & 0.0222 & 0.0220 & 20344.13 & 19990.00 & 20152.63 & 4.114$\times10^{8}$ \\

Fe &    5.0 & 0.4389 & 0.4200 &   147.00 &    80.00 &   145.63 & 4.827$\times10^{5}$ \\
Fe &   10.0 & 0.3084 & 0.3000 &   346.62 &   150.00 &   343.26 & 1.374$\times10^{6}$ \\
Fe &  100.0 & 0.1062 & 0.1070 &  3128.82 &  1800.00 &  3109.18 & 3.989$\times10^{7}$ \\
Fe &  500.0 & 0.0494 & 0.0500 & 20082.96 & 14100.00 & 20002.83 & 4.289$\times10^{8}$ \\
Fe & 1000.0 & 0.0351 & 0.0350 & 43303.76 & 32200.00 & 43155.53 & 1.209$\times10^{9}$ \\

\hline
\end{tabular}

\caption{Structural and cohesive properties of magnetized 1D atomic chains for hydrogen (H), helium (He), carbon (C), and iron (Fe) under magnetic fields $B_{12}$ (in units of $10^{12}$ G). The table lists equilibrium lattice constants ($a_{eq}$), uncorrected cohesive energies ($Q_s^0$), corrected cohesive energies ($Q_s$), and mass densities ($\rho$). Reference values from previous studies \cite{medin2008thesis} are included for comparison.}
\label{tab:structural}
\end{table*}

Table~\ref{tab:structural} summarizes the calculated structural and cohesive properties of magnetized one-dimensional atomic chains of H, He, C, and Fe over a broad range of magnetic field strengths. As the magnetic field increases, the equilibrium lattice spacing decreases substantially due to enhanced transverse magnetic confinement, leading to highly compressed chain-like structures. This compression is accompanied by a rapid increase in the cohesive energies and mass densities, reflecting the progressive stabilization of condensed matter under extreme magnetic fields.
This behavior directly reflects the transverse 
compression induced by Landau quantization, which effectively squeezes the 
electronic cloud into narrow cylindrical structures aligned along the magnetic 
field direction, thereby enhancing the electron-nucleus attraction along the 
longitudinal axis.
The effect is particularly pronounced for heavier elements such as carbon and iron, which exhibit cohesive energies reaching tens of keV at the highest field strengths considered.

The calculated equilibrium lattice constants show excellent agreement with previously reported values across all magnetic fields, with deviations typically below a few percent. The cohesive energies also follow the same qualitative trends as the reference calculations, although systematic quantitative differences are observed, especially for iron at high magnetic fields. These discrepancies primarily arise from differences in computational methodology. In the present work, isolated atomic energies are computed using the same real-space supercell framework employed for the periodic chain calculations, ensuring internal consistency between the atomic and condensed phases. In contrast, the reference calculations of Medin \& Lai  employed separate atomic and condensed-matter codes optimized independently for each problem. Since the cohesive energy is obtained as a small difference between two very large total energies in the strongly magnetized regime, even modest methodological differences in the treatment of the atomic reference state can lead to appreciable variations in the final cohesive energies.

Taken together, the results demonstrate that the present implementation of \codename\ reliably reproduces previously reported neutron-star condensed matter calculations over a wide range of field strengths and atomic species, while also providing a robust platform for exploring regimes where occupation complexity and extreme magnetic anisotropy make conventional electronic-structure calculations particularly challenging.

\begin{table*}
\centering
\small
\setlength{\tabcolsep}{4pt}

\begin{tabular}{l r r r r r r r}
\hline
El & $B_{12}$ & $k$ & $\gamma$ & $v_s$ & $\Theta_D$ & $T_{melt}$ & Phase \\
 & ($10^{12}$G) & (eV/bohr$^2$) &  & ($c$) & (K) & (K) & \\
\hline

H  &    1.0 & 3.277$\times10^{3}$ & 2.995 & 0.043 & 2.540$\times10^{5}$ & 2.010$\times10^{4}$ & LIQUID \\
H  &   10.0 & 4.985$\times10^{4}$ & 2.772 & 0.067 & 9.906$\times10^{5}$ & 4.890$\times10^{4}$ & LIQUID \\
H  &  100.0 & 7.478$\times10^{5}$ & 2.901 & 0.105 & 3.837$\times10^{6}$ & 1.208$\times10^{5}$ & LIQUID \\
H  & 1000.0 & 1.006$\times10^{7}$ & 4.101 & 0.160 & 1.407$\times10^{7}$ & 2.790$\times10^{5}$ & LIQUID \\

He &    1.0 & 6.061$\times10^{3}$ & 2.728 & 0.036 & 1.733$\times10^{5}$ & 5.527$\times10^{4}$ & LIQUID \\
He &   10.0 & 1.121$\times10^{5}$ & 2.817 & 0.060 & 7.455$\times10^{5}$ & 1.554$\times10^{5}$ & LIQUID \\
He &  100.0 & 1.816$\times10^{6}$ & 2.935 & 0.096 & 3.000$\times10^{6}$ & 3.992$\times10^{5}$ & LIQUID \\
He & 1000.0 & 2.745$\times10^{7}$ & 2.698 & 0.151 & 1.166$\times10^{7}$ & 9.896$\times10^{5}$ & LIQUID \\

C  &    1.0 & 4.197$\times10^{3}$ & 5.220 & 0.030 & 8.327$\times10^{4}$ & 1.181$\times10^{5}$ & LIQUID \\
C  &   10.0 & 1.848$\times10^{5}$ & 5.328 & 0.063 & 5.525$\times10^{5}$ & 5.145$\times10^{5}$ & LIQUID \\
C  &  100.0 & 6.686$\times10^{6}$ & 2.923 & 0.138 & $\mathbf{3.323\times10^{6}}$ & $\mathbf{2.466\times10^{6}}$ & SOLID \\
C  & 1000.0 & 1.180$\times10^{8}$ & 2.728 & 0.228 & $\mathbf{1.396\times10^{7}}$ & $\mathbf{6.722\times10^{6}}$ & SOLID \\

Fe &    5.0 & 2.793$\times10^{4}$ & 3.176 & 0.032 & 9.962$\times10^{4}$ & 6.245$\times10^{5}$ & LIQUID \\
Fe &   10.0 & 1.689$\times10^{5}$ & 3.587 & 0.056 & $\mathbf{2.449\times10^{5}}$ & $\mathbf{1.864\times10^{6}}$ & SOLID \\
Fe &  100.0 & 5.775$\times10^{6}$ & 4.288 & 0.112 & $\mathbf{1.432\times10^{6}}$ & $\mathbf{7.562\times10^{6}}$ & SOLID \\
Fe &  500.0 & 9.607$\times10^{7}$ & 3.496 & 0.212 & $\mathbf{5.842\times10^{6}}$ & $\mathbf{2.720\times10^{7}}$ & SOLID \\
Fe & 1000.0 & 3.288$\times10^{8}$ & 4.552 & 0.279 & $\mathbf{1.081\times10^{7}}$ & $\mathbf{4.688\times10^{7}}$ & SOLID \\

\hline
\end{tabular}

\caption{Elastic and thermal properties of magnetized 1D atomic chains. The table summarizes the effective spring constant ($k$), Gr\"uneisen parameter ($\gamma$), longitudinal sound speed ($v_s$), Debye temperature ($\Theta_D$), and melting temperature ($T_{melt}$). Boldface values correspond to systems identified to be in the solid phase according to the Lindemann-like stability criterion, while non-bold entries correspond to liquid/gaseous phases.}
\label{tab:thermal}
\end{table*}


\begin{figure*}[t]
    \centering
    \includegraphics[width=0.92\textwidth]{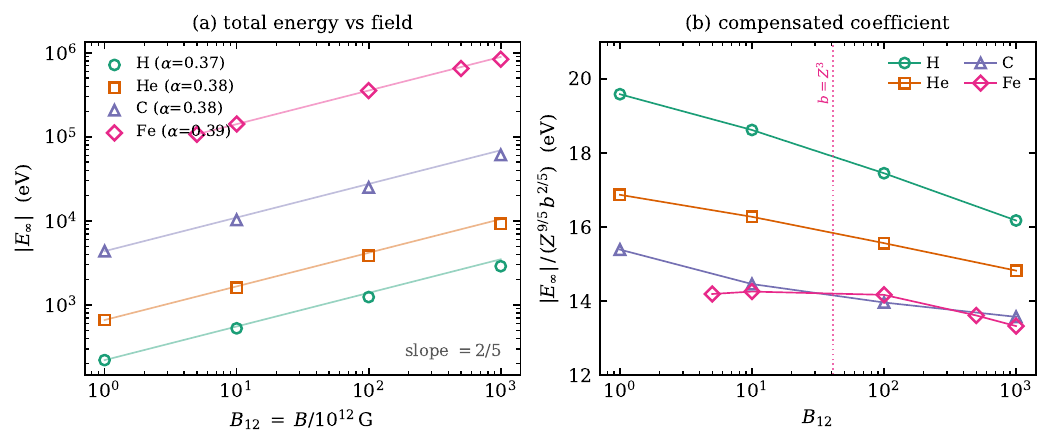}
    \caption{Scaling of the \codename\ total energy per cell with magnetic
    field strength. (a) $|E_\infty|$ versus $B_{12}$ on logarithmic axes for
    H, He, C and Fe; symbols are computed values (FFT solver), grey lines are
    guides of slope $2/5$, and fitted exponents $\alpha$ (from
    $|E_\infty|\propto B_{12}^{\alpha}$) are listed in the legend.
    (b) Compensated coefficient $|E_\infty|/(Z^{9/5}b^{2/5})$; its
    near-constancy confirms the Landau--Thomas--Fermi law of
    Eq.~(\ref{eq:scaling}). The dotted line marks the crossover $b=Z^3$
    ($B_{12}\approx41$ for iron) to the lowest-Landau-level regime.}
    \label{fig:scaling}
\end{figure*}

\subsection{Scaling behaviour and physical consistency}
A stringent, parameter-free check of the solver is provided by the known
asymptotic scaling of the total energy with magnetic-field strength. It is
convenient to introduce the dimensionless field $b=B/B_0$, where
$B_0=m_e^2e^3c/\hbar^3=2.35\times10^{9}$~G is the field at which the electron
cyclotron energy equals twice the Rydberg; in our units $b\simeq425\,B_{12}$.
For an atom or chain of nuclear charge $Z$, three regimes are established
\cite{Kadomtsev1970,MuellerRauSpruch1971,Lieb1994,Lai2001}: a weak-field
Thomas--Fermi regime ($b\ll Z^{4/3}$, $|E|\propto Z^{7/3}$); an intermediate
Landau--Thomas--Fermi regime ($Z^{4/3}\ll b\ll Z^3$) in which the binding
energy follows
\begin{equation}
   |E|\;\propto\;Z^{9/5}\,b^{2/5},
   \label{eq:scaling}
\end{equation}
and a fully spin-polarized lowest-Landau-level regime ($b\gg Z^3$), where
$|E|\propto Z^{3}\,[\ln(b/Z^3)]^2$. The condensed-matter analogues of these
laws follow from magnetized Thomas--Fermi theory \cite{Fushiki1989}.

Figure~\ref{fig:scaling}(a) shows the \codename\ total energies per cell for
H, He, C and Fe over three decades in field. A power-law fit
$|E_\infty|\propto B_{12}^{\alpha}$ gives $\alpha=0.37$--$0.39$ for all four
species, in close agreement with the Landau--Thomas--Fermi exponent $2/5$ of
Eq.~(\ref{eq:scaling}) (grey guide lines). The agreement is made quantitative
in Fig.~\ref{fig:scaling}(b), where the compensated coefficient
$|E_\infty|/(Z^{9/5}b^{2/5})$ is nearly constant: for iron it stays within
$\sim5\%$ of $14$~eV across the entire range $B_{12}=5$--$1000$. The mild
downward drift at the highest fields is itself expected --- the crossover to
the lowest-Landau-level regime occurs at $b=Z^3$, i.e. $B_{12}\approx41$ for
iron (dotted line), beyond which the growth slows towards the logarithmic
$[\ln(b/Z^3)]^2$ form. Carbon, helium and hydrogen have their crossover at
$B_{12}\ll1$ and thus lie entirely within the lowest-Landau-level regime,
consistent with the slightly stronger softening of their compensated
coefficients. That the solver reproduces both the correct exponent and the
correct prefactor, without any fitting, provides strong evidence that the
total energies are physically sound even where direct numerical benchmarks are
sparse.

This analysis also clarifies the trends in Table~\ref{tab:benchmarks}. Whereas
the total energy grows as $B_{12}^{2/5}$, the Fermi energy is a surface
property of the band structure and grows far more slowly --- empirically
$|\epsilon_F|\propto B_{12}^{1/4}$ for iron --- and is governed by the dense
manifold of weakly occupied, nearly degenerate $m$-orbitals near $\epsilon_F$
(cf. Fig.~\ref{fig:occupation}). The increasing deviations of $\epsilon_F$ at
the highest fields therefore reflect the intrinsic numerical difficulty of
locating a single global Fermi level within this manifold, rather than a
failure of the underlying energetics, which Eq.~(\ref{eq:scaling}) confirms
remain accurate.

\subsection{Macroscopic properties}
Having validated the accuracy of the present framework against previously reported structural and cohesive properties, we now turn to the investigation of macroscopic quantities that extend beyond the scope of earlier calculations. Table~\ref{tab:thermal} summarizes the calculated elastic and vibrational properties of strongly magnetized one-dimensional atomic chains of H, He, C, and Fe over a wide range of magnetic field strengths. Because the intense magnetic field strongly suppresses transverse electronic motion, the vibrational dynamics are dominated by longitudinal lattice oscillations propagating along the chain direction.

The effective spring constant $k$, obtained from the harmonic curvature of the total-energy surface, increases rapidly with magnetic field strength for all elements considered. This behavior reflects the increasing rigidity of the magnetically confined lattice as enhanced electronic localization strengthens the longitudinal restoring forces between neighboring ions. Consequently, both the longitudinal sound speed $v_s$ and the Debye temperature $\Theta_D$ exhibit a strong systematic increase with increasing magnetic field. At the highest fields, the sound speeds become a substantial fraction of the speed of light, particularly for heavier elements. For example, the iron chain at $B_{12}=1000$ attains a longitudinal sound velocity of approximately $0.279~c$, indicating an extremely stiff quasi-one-dimensional lattice. The corresponding Debye temperatures reach values approaching $10^7$~K, implying that quantum lattice vibrations and zero-point motion remain important even at typical neutron-star surface temperatures.

The calculated Gr\"uneisen parameters further illustrate the strong coupling between lattice compression and vibrational behavior in these systems. Although $\gamma$ generally remains of order unity, significant variations occur across different elements and magnetic field strengths. The relatively large values observed in several cases suggest appreciable anharmonicity in the effective lattice potential. Such behavior is expected in highly anisotropic magnetized matter, where the longitudinal and transverse degrees of freedom respond very differently to compression and electronic confinement.

The calculated melting temperatures and phase classifications reveal a pronounced contrast between light and heavy elements. For hydrogen and helium, the calculated melting temperatures remain far below the characteristic magnetar surface temperature of approximately $10^6$~K. Furthermore, the large zero-point vibrational energy in these light systems significantly disrupts cohesive binding, precluding the formation of long-range ordered configurations across the investigated magnetic-field range. We emphasize, however, that under such extreme magnetic and thermodynamic conditions, the distinction between dense liquids, supercritical fluids, and partially ionized gaseous phases becomes intrinsically ambiguous. The present classification should therefore be interpreted primarily as a measure of lattice stability rather than as a strict thermodynamic phase boundary.

In contrast, heavier elements possess sufficiently large ionic masses and stronger electrostatic binding to stabilize condensed structures against quantum fluctuations. While carbon and iron behave as liquid-like chains at lower magnetic fields, both undergo a transition to stable solid phases at higher field strengths. Carbon becomes mechanically stable for $B_{12}\ge100$, whereas iron exhibits solid behavior already at $B_{12}=10$. At still higher fields, the thermal stability becomes extraordinary. For example, the iron chain at $B_{12}=500$ exhibits a Debye temperature of approximately $5.8\times10^6$~K and a melting temperature approaching $2.7\times10^7$~K, demonstrating that condensed neutron-star matter can remain mechanically robust even under extreme astrophysical conditions.

Overall, these results demonstrate the profound restructuring of condensed matter in ultra-strong magnetic fields. Increasing magnetic confinement simultaneously compresses the lattice, enhances the elastic stiffness, raises the characteristic vibrational energy scales, and promotes solidification in heavier elements. The progression from weakly bound, liquid-like hydrogen chains to ultra-dense and highly rigid iron structures highlights the emergence of a fundamentally distinct condensed matter regime governed by magnetic quantization, anisotropic confinement, and quasi-1D bonding.

\section{Conclusion}
In this work, we have presented a robust real-space density functional theory 
framework, \codename, specifically designed to simulate periodic 
one-dimensional atomic chains in the ultra-strong magnetic fields characteristic 
of magnetar environments. By utilizing a grid-based linear fast Fourier 
transform convolution approach, our solver successfully overcomes the numerical 
instabilities and artificial band-crossing challenges that frequently plague 
standard point-by-point integration methods in highly anisotropic, 
magnetically confined regimes.

The code is tested for various atomic chains at various strengths of magnetic fields from 1$\times 10^{12}$ G  to $1000 \times 10^{12}$ G. The 
framework is capable of cleanly resolving a highly complex, degenerate manifold of states 
near the Fermi level. \codename\ is capable of generating detailed electronic structure including band structure, DOS, 3D density, fractional occupations, etc. 

Systematic evaluation of the thermodynamic and macroscopic properties across a 
broad range of magnetic field strengths  reveals the 
profound restructuring of condensed matter in these extreme astrophysical 
regimes. The extreme transverse confinement induced by Landau quantization 
dramatically compresses the lattice constants and drives macroscopic mass 
densities up by several orders of magnitude. This structural compression is 
accompanied by a substantial elastic stiffening of the chains, yielding 
extraordinarily high Debye temperatures and longitudinal sound speeds that 
approach a significant fraction of the speed of light.

Furthermore, our thermal stability analysis provides crucial insights into the 
composition of magnetar crusts. Under a standard baseline surface temperature 
of $10^{6}$~K, light elements like Hydrogen and Helium remain in a highly mobile, 
non-crystalline state. We classify these as liquid-like, with the physical 
caveat that they likely behave as supercritical fluids or dense, highly ionized 
gaseous plasmas lacking long-range order. In contrast, heavier elements like 
Carbon and Iron easily overcome zero-point fluctuations due to their larger 
nuclear charges and masses, undergoing clear transitions to highly rigid, solid 
polymeric phases. The remarkably high melting temperatures calculated for Iron 
at $B_{12} = 500$ confirm that the outer crusts of strongly magnetized 
neutron stars will maintain a robust, solid crystalline structure even under 
intense ambient thermal conditions.

Ultimately the \codename\ offers a robust method to compute the essential condensed matter properties of matter under extreme magnetic field. We believe it is an essential tool to understand ever increasing data from modern sensors.  By making the source code available to wider scientific community, we believe the code can be scrutinized, tested, and subsequently grow by community contribution.

\section{Code availability}
 \codename, is available under Creative Commons NC-SA license and can be downloaded from:\\ \codeurl.\\ The code is also accompanied by additional resources to generate the data used in this manuscript. 

\section*{Acknowledgments}
D.M. acknowledges the support of the Department of Atomic Energy, Government of India, under project No. 12-R\&D-TFR-5.02-0700. B.P. thanks Prof. D. G. Kanhere for discussion. 
\printcredits

\appendix
\section{Example configuration file}
\label{sec:config}
Below is an example configuration file. The code requires the name of the file to be \texttt{config.yaml}. 
\begin{lstlisting}
grid:
  Nk: 50
  Nz: 901

output:
  create_basic_plots: true
  save_data: true

scf:
  max_iter: 120
  mixing: 0.25
  tol: 0.0001

simulation:
  B_field: 500.0e+12
  M_max: 160
  atomic_Z: 26
  lattice_constant: 0.05
  nu_max: 2
  num_electrons: 26
  use_fft: true
\end{lstlisting}

\bibliographystyle{cas-model2-names}

\bibliography{cas-refs}


\end{document}